 \documentstyle[prl,aps,epsf]{revtex}
\include{psfig} %\usepackage{graphicx}

\begin{document}

\draft

\wideabs{
\title{ Andreev Level Qubit }
\author{ A.\ Zazunov$^a$, V.\ S.\ Shumeiko$^a$, E.\ N.\ Bratus'$^b$,
J.\ Lantz$^a$,
and G.\ Wendin$^a$}
\address{$^a$ Department of Microelectronics and Nanoscience,
Chalmers University of Technology and
G\"{o}teborg University, \\S-41296 G\"{o}teborg, Sweden \\
$^b$ B.\ Verkin Institute for Low Temperature Physics and
Engineering
\\310164 Kharkov, Ukraine}
%

%\date{\today}
\maketitle

\begin{abstract}
We investigate the  dynamics of a two-level Andreev bound state 
system in a transmissive quantum point contact embedded in an rf-SQUID. 
Coherent coupling of the Andreev levels to the circulating supercurrent 
allows manipulation and read out of the level states. The two-level 
Hamiltonian for the Andreev levels is derived, and the effect of 
interaction with the quantum fluctuations of the induced flux is studied.
We also consider an inductive coupling of qubits, and discuss the 
relevant SQUID parameters for qubit operation and read out. 
\end{abstract}
}

\narrowtext
%\twocolumn

%%%%%%%%%%%%%%%%%%%%
Recent observations of quantum coherence in 
superconducting circuits\cite{Nakamura,Friedman,Mooij2,Vion,MQT}  
have made superconducting qubits a realistic possibility. 
Superconducting qubits employ the phenomenon of 
macroscopic quantum coherence (MQC)\cite{Leggett1984}, and operate with 
coherent superpositions of quantum states of a macroscopic object - a
superconducting condensate. An elementary MQC circuit consists
of a hysteretic SQUID with a small capacitance Josephson tunnel junction
(persistent current or flux qubit). The qubit is biased at half-integer
external flux, and operates with the two degenerate current states
corresponding to the clockwise and counter-clockwise circulating
persistent currents. The coupling of the current states is due to 
quantum fluctuations of the electric charge on the junction capacitor. 
Quantum measurements of the fluctuating persistent current, or induced 
magnetic flux, provide means to read out the qubit state. Further 
modifications of the flux qubits involve implementation of multiple 
junction circuits \cite{Mooij2,Vion}.
%(for a review, see Makhlin et al. \cite{Makhlin2001}).
In multiple junction MQC circuits, it is possible to employ one 
of the available dynamic variables for qubit operation, and another one
for qubit readout\cite{Vion,Zorin2002}. 

In this paper we consider a new type of superconducting qubit where the
switching between the two persistent current states in a SQUID is 
achieved by employing a true {\em microscopic} system formed by the 
two-level Andreev bound 
states in a superconducting atomic-size quantum point contact embedded 
in the SQUID \cite{Jonn2002}. In this Andreev level qubit, the quantum 
information is stored in the microscopic quantum system, the Andreev bound 
states, similar to non-superconducting solid state qubits like localized spins 
on impurities \cite{Kane} or  quantum dots\cite{Loss}. 
Read-out of the Andreev level qubit is achieved by 
monitoring the macroscopic persistent
current or the induced flux in the SQUID, similar to the MQC qubits.

The Josephson effect in a single atomic-size quantum point contact (QPC) 
embedded in a low-inductance non-hysteretic SQUID (Fig. 1)
has first been investigated by Koops et al.\cite{Koops1996}; 
in this experiment, the averaged current-phase relation in the ground 
state was measured by performing a classical measurement 
of the induced flux. The results of this experiment, and also of other
experiments on atomic-size QPCs where the critical current\cite{Goffman}
and current-voltage characteristics\cite{Post,Scheer} have been
investigated, are consistent with a theoretical picture in which 
the Andreev bound levels play the central role in the
Josephson current transport\cite{Furusaki1991}.
\begin{figure}[t]
\centerline{\psfig{figure=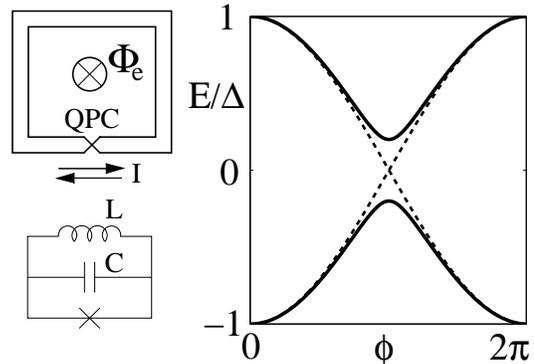,width=0.8\linewidth}}
\label{fig:alqubit}

\caption{Left: Sketch of the Andreev level qubit - a non-hysteretic
rf-SQUID with a quantum point contact (QPC), and the equivalent circuit
containing the Josephson junction and $LC$-oscillator. Right: The energy
spectrum of the QPC with finite reflectivity ($R=0.04$) (solid line),
appears as a hybridization of the current states ($R=0$) (dashed line).
}
\end{figure}

The Andreev bound levels are formed in a QPC due to Andreev reflections
by the discontinuity of the superconducting phase at the contact in the
presence of the applied current. The Andreev level wave functions are
localized in the vicinity of the contact over a distance of the order of
the superconducting coherence length, and the number of Andreev bound
levels is limited to one pair of levels per conducting electronic
mode.  Thus, a superconducting QPC may be viewed as a kind of quantum dot
which contains a finite number of localized quantum states. In highly
transmissive QPCs, the Andreev levels lie deep within the superconducting
gap and are well decoupled from the continuum quasiparticle states in the
electrodes.

The most important property for qubit applications 
is the coherent coupling of the Andreev levels to the supercurrent 
flowing through the contact.
This makes the Andreev levels accessible for manipulations and for
measurements. Extensive studies have shown that the time evolution of 
the Andreev levels can be controlled by applying resonant rf flux pulses 
\cite{Shumeiko1993} or by ramping the external
flux\cite{Jonn2002}. By these means, one can drive
Andreev levels out of the ground state and prepare any excited
state\cite{Moriond}. On the other hand, the time evolution of the Andreev levels 
changes the current through the QPC. Thus the QPC operates as a quantum 
switch which controls the direction of the circulating current in the SQUID. 
Performing quantum measurement of 
the circulating current, or of the corresponding induced flux, in the SQUID
one is able to measure the state of the Andreev level system. 
Fidelity of such a measurement
requires the qubit evolution to be slow on a time scale of intrinsic
electromagnetic fluctuations in the SQUID. Then the current 
(and induced flux) averaged over electromagnetic fluctuations will
adiabatically follow the qubit evolution.

It is important to note that the interaction of the Andreev levels with electromagnetic
modes in the SQUID is an essential element of the qubit dynamics: 
To maintain the current switching, the plasma
frequency of the SQUID must be sufficiently large.
Furtheremore, since the current in the 
single-mode QPC undergoes strong quantum fluctuations, the induced flux 
and hence the superconducting phase difference, are also fluctuating 
quantities, and the 
theory of the Andreev level qubit should include a full
quantum mechanical treatment of the coupled Andreev levels and
electromagnetic fluctuations.

To derive an effective quantum Hamiltonian describing coupled
Andreev levels
and electromagnetic fluctuations, we employ a path integral approach
commonly used in MQC theory\cite{Ambegaokar1982}. The central problem
here is to extend the theory, originally developed for tunnel
junctions, to the for us interesting case of {\em high transmission} QPCs;
this problem is solved by using the exact boundary condition in the action
instead of the tunnel Hamiltonian.  Following Ref.
\cite{Ambegaokar1982}, we present the evolution operator for the
system on the form,
\begin{equation}
U = \int {\cal D}\phi \; e ^{iS_{osc}[\phi]/\hbar}
\int {\cal D}^2\psi_L {\cal D}^2\psi_R \;
e^{iS_J/\hbar},
\label{U}
\end{equation}
where the integration is performed over the superconducting phase
difference $\phi(t)$ and Nambu-Grassman fields $\psi_{L,R}(r,t)$
representing electronic degrees of freedom in the left and right
electrodes respectively. The action $S_{osc}[\phi]$ describes an
$LC$-oscillator formed by the junction capacitance $C$ and the
superconducting loop inductance $L$ (see Fig. 1), while the action $S_J$ describes 
the Josephson junction.
%The Lagrangian of the $LC$-oscillator has the form
%${\cal L}_{osc} = (C\hbar^2/8e^2)(\partial_t\phi)^2 -
%(c\hbar/2e)^2(\phi^2/2L)$,
%where $\phi$ is related to the induced flux, $\Phi=(c\hbar/2e)\phi$.

To derive the action for the Josephson junction, we consider the two 
superconducting reservoirs coupled via a single-channel localized 
scatterer. The scatterer is represented by a normal electron scattering
matrix, which imposes the boundary condition for the quasiclassical wave 
functions of the quasiparticles in the reservoirs. The microscopic 
Hamiltonian for the reservoirs,
\begin{equation}
H_{} = \sum_{\sigma=L,R}\int dr \;\psi^\dagger_{\sigma} \hat h
\psi_{\sigma},\;
\hat h = \left({\hat p^2\over 2m}-\mu\right)\sigma_z + \Delta\sigma_x ,
\label{HLR}
\end{equation}
is considered within the mean field approximation, the superconducting
phase being gauged out and included in the boundary condition.
The Hamiltonian generates the Lagrangian,
$ {\cal L}= \sum_{\sigma=L,R}\int dr\; \bar\psi_{\sigma}
(i\hbar\partial_t -\hat h ) \psi_{\sigma}$.
 %$\sigma_i$ are the Pauli matrices in the Nambu space.
In the boundary condition, which connects the fields $\psi_{L,R}(0,t)$ 
at the contact, the energy dependence of the scattering matrix on the
scale of 
$\Delta$ can be neglected, so that the same matrix describes both the 
electrons and holes.  Furthermore, without loss of generality, it is
possible to eliminate constant scattering phases from the boundary
condition and include them in the positive and 
negative momentum components of the fields $\psi_\sigma(r,t)$. 
Within the quasiclassical approximation, such a transformation will not 
affect the Hamiltonian in Eq.
(\ref{HLR}). The boundary condition can then be written on the form,
$
\chi(t) =  de^{-i\sigma_z \phi(t)/2}\psi_L(0,t) - (1+r)\psi_R(0,t) =0,
$
where $d=\sqrt D$ and $r=\sqrt R$ are real transmission and reflection
amplitudes respectively.
This boundary condition is incorporated into the junction action $S_J$
by means of the Lagrange field $\eta(t)$,
\begin{equation}
e^{iS_{J}/\hbar} =
\int {\cal D}^2 \eta \; e^{ (i/\hbar)\int dt ({\cal L} - \bar \eta\chi -
\bar\chi \eta )}.
\label{}
\end{equation}
We will see later that the fermionic field $\eta(t)$ describes the
Andreev levels.

Integration over the rapidly varying fermionic fields in the electrodes,
$\psi_\sigma(r,t)$, yields the effective action
\begin{equation}
e^{iS_{eff}[\phi]/\hbar} = \int{\cal D}^2\eta \; e^{(i/\hbar)\int dt_1
dt_2 \;
\bar\eta(t_1)G(t_1, t_2) \eta(t_2)}  ,
\label{SJ}
\end{equation}
where
\begin{eqnarray}
G(t_1,t_2) = 
-{D \over (1+r)^2} e^{-i\sigma_z\phi(t_1)/2} g(t_1 - t_2)
e^{i\sigma_z\phi(t_2)/2} \nonumber\\
- g(t_1-t_2), \;\;\;\;\;\;\;\;g(\omega) = \sum_{p} (\hbar\omega - \hat h)^{-1}.
\label{G12}
\end{eqnarray}
Equation (\ref{G12})
provides the required generalization of the 
the effective action of the tunnel theory\cite{Ambegaokar1982} 
to junctions with {\em arbitrary} transparency.
In the low frequency limit, $\omega \ll \Delta/\hbar$, the Green's
function $g(\omega)$ reduces to a simple form,
$g(\omega) = (-\pi\nu_F/\Delta) (\hbar\omega +\Delta\sigma_x)$
($\nu_F$ is the density of states at the Fermi level),
which leads to the local-in-time effective Lagrangian, $ G(t_1, t_2)
=G(t_1)\delta(t_1-t_2)$,
\begin{equation}
G(t) = i\hbar\partial_t - {1-r\over 4}\;\hbar\partial_t\phi\sigma_z +
\left(
\begin{array}{ll}
 0& z e^{-i\phi/2}\\
z^\ast e^{i\phi/2} &0\\
\end{array}\right),
\label{G}
\end{equation}
with $z=\Delta [\cos(\phi/2) + ir\sin(\phi/2)].$

Equation  (\ref{G}) describes a two-level fermionic system. 
%controlled by a classical variable $\phi(t)$. 
Under stationary conditions, $\partial_t\phi=0$,
the spectrum of the system,  $\hbar\omega = \pm E_a$,
$
E_a=\Delta\sqrt{\cos^2(\phi/ 2)+R\sin^2(\phi/ 2)},
$
coincides with the Andreev level spectrum \cite{Furusaki1991},
shown in Fig. 1. It follows from this equation
that the assumed low-frequency approximation is
appropriate for transparent contacts,  $R\ll 1$, at $\phi\approx\pi$,
where the Andreev level energy is small, $E_a\ll \Delta$.
To derive the Hamiltonian for the Andreev two-level system,
it is convenient first to eliminate the time derivative of the phase in
Eq. (\ref{G}) by means of a unitary rotation, and then to use the
relation, $G(t)  = i\hbar \partial_t - \hat H_{a}$, to obtain
\begin{equation}
\hat H_{a}= \left(
\begin{array}{cc}
        0 & - z e^{-ir\phi/2} \\
        - z^\ast e^{ir\phi/2} & 0
\end{array}
\right).
\label{Ha}
\end{equation}
This equation is the first main result of the present paper \cite{Ha}.

The slow dynamics in transparent contacts, $R\ll 1$, is described
by two variables, $\eta(t)$ and $\phi(t)$. 
This is rather different from tunnel contacts with $D\ll 1$, where the 
effective action only depends on the phase difference and basically reduces
to the potential Josephson energy \cite{Ambegaokar1982}. 
This difference is easily understood if one
takes into account that the Andreev levels in tunnel junctions
are close to the gap edge, $E_a\approx \Delta$, making $\eta(t)$ a rapid
variable. Integration  over $\eta$ in Eq. (\ref{SJ}) assuming small $D$ in 
Eq. (\ref{G12}), recovers the effective action of Ref.\cite{Ambegaokar1982}.

Let us now consider the current through the junction. 
The statistically averaged current $\langle I\rangle $ can be expressed
through the Josephson part, $U_J$, of the evolution operator in Eq.
(\ref{U}),
$
\langle I\rangle = 2ei (\delta / \delta\phi)\ln \,\langle U_J
\rangle .
$
In terms of the effective action, the equation for the averaged
current  reduces to the form $\langle I\rangle =
Tr(\rho_\eta\hat I) $, where $\rho_\eta$ is the density
matrix, and $\hat I$ is the current operator of the two-level Andreev
system,
\begin{equation}
\hat I = {2e\over\hbar}\;{d\hat H_{a}\over d\phi}
=  {e{\cal I}(\phi) \over\hbar} \left(
\begin{array}{ll}
        0 & e^{-ir\phi/2} \\
        e^{i r\phi/2} & 0
\end{array}
\right),
\label{}
\end{equation}
${\cal I}(\phi) = \Delta D \sin(\phi/2)$.
The current operator $\hat I$ does not commute with the Hamiltonian 
$\hat H_{a}$,
which is a consequence of the normal electron reflection at the QPC.
Therefore the Andreev levels consist of superpositions of the current
eigenstates, unless $R=0$, and hence the current expectation value in
the Andreev state, $I_a =
(2e/\hbar)(dE_a/d\phi)=(e/2\hbar)(D\Delta^2/E_a)\sin\phi$, does not
coincide with the current eigenvalues, $\pm e{\cal I}/\hbar$, which are
evaluated during the quantum measurements.
Furthermore, the Andreev level current undergoes quantum fluctuations
with the spectral function (cf. Ref.\onlinecite{Rodero}),
$
S(\omega) =  I_a^2 R\tan^2(\phi/2)\delta(\omega-2E_a).
$
In the SQUID geometry, these fluctuations generate strong quantum
fluctuations of the phase.

We now take the quantum dynamics of the superconducting phase
into the consideration.
The quantum Hamiltonian of the $LC$-oscillator associated with the action
${S}_{osc}[\phi]$ in Eq. (\ref{U}) has the form,
$
\hat H_{osc} = -(\hbar \partial_\phi)^2/2M + M\omega^2\tilde\phi^2/2,
 $
where $M=\hbar^2/8E_C$, $\omega= \sqrt{8E_LE_C/\hbar^2}$, 
$E_C=e^2/2C$, $E_L =(\hbar c/2e)^2(1/L)$, and   
$\tilde\phi$ is related to the induced flux, $\tilde\Phi=(c\hbar/2e)\tilde\phi$.

In the practically important case of small loop inductance,
$E_J \ll E_L $, the SQUID is in the non-hysteretic
regime, and the induced flux is small, $\tilde\phi\ll 1$. Introducing
the phase difference $\phi_e$ related to a stationary external
flux we expand the Hamiltonian in Eq. (\ref{Ha}) over small $\tilde\phi$; then
the Hamiltonian of the whole system in the current eigenbasis takes the form,
\begin{equation}
\hat H= -\Delta\left(
\cos{\phi_e\over 2}\;\sigma_z +
r \sin{\phi_e\over 2}\;\sigma_x
\right) +
  {{\cal I}(\phi_e) \over 2}\;\tilde\phi\sigma_z + \hat H_{osc}
  %{r\dot\phi_e\over 2}\sigma_z +
  \label{Hlin}
  \end{equation}
  %
 %$$ \frac{\hat p^2}{2M} + {M\omega^2\over 2} \phi^2. $$
This Hamiltonian describes a spin degree of freedom linearly coupled
to an oscillator, the steady state of the oscillator being shifted
from the origin by $ \pm {\cal I}/2M\omega^2$ depending on the spin
direction (direction of the current in the junction). We are interested
in the case when the induced flux adiabatically follows the evolution of 
the Andreev levels. This regime
corresponds to a large oscillator frequency compared to the
Andreev level spacing, $\hbar\omega \gg 2E_a$. In this case,
the oscillator can be assumed to be in the ground state,
$\varphi_0(\phi_\pm) $, $ \phi_\pm = \tilde\phi \pm{\cal I}/2M\omega^2$,
since the
probability of transitions among the oscillator levels is small.
Averaging out the ground state phase fluctuations \cite{Leggett1987},
we finally arrive at the effective Hamiltonian describing the Andreev
level qubit,
\begin{equation}
  \hat H_{q} =
-\Delta\left(
\cos{\phi_e\over 2}\;\sigma_z +
q_0 r \sin{\phi_e\over 2}\;\sigma_x\right).
 %+{r\dot\phi_e\over 2}\sigma_z .
  \label{Hq}
\end{equation}
The factor, $ q_0 = \exp \left(-{\cal I}^2/ 4M\hbar\omega^3\right)$,
is the overlap integral between the oscillator ground state wave
functions for different current directions, $\varphi_0(\phi_\pm)$.
The averaged value of the induced phase is then given by
$\langle \tilde\phi\rangle = ({\cal I}/2M\omega^2)Tr(\rho_\eta\sigma_z) $,
which
yields the relation between the induced flux operator and the current
operator, $\hat{\tilde\Phi} = (L/c)\hat I $. Therefore quantum measurement of
the flux on a time scale $1/\omega<\tau<\hbar/2E_a$ will allow
correct evaluation of the Andreev level state.

The qubit Hamiltonian (\ref{Hq}) is another main result of this
paper. Equation  (\ref{Hq}) is equivalent to the Hamiltonian
of non-interacting Andreev levels, Eq. (\ref{Hlin}), but with
reduced reflectivity, $\tilde R =  q_0^2 R$. 
One may interpret this reduction as the effect of the inertia of the loop 
oscillator, which makes it more difficult for the Andreev levels to switch 
direction of the current. The effect becomes increasingly strong in the 
limit of a classical oscillator with large ''mass''. This renomalization 
effect leads to reduction of the Andreev
level energy $E_a\rightarrow\tilde E_a$, and hence to the reduction
of the frequency of the qubit rotation, remaning the results of 
Ref.\cite{Shumeiko1993,Moriond} essentially the same. This might
be important for practical applications, because it would allow the 
frequency of the qubit rotation to be tuned by choosing circuit parameters 
rather than by tuning the contact reflectivity.

The following chain of inequalities summarizes the requirements
for the Andreev level qubits,
\begin{equation}
 2\Delta\sqrt {\tilde R}  \ll \hbar\omega \ll\Delta\sim E_J \ll E_L.
\label{}
\end{equation}
These requirements are consistent with the typical circuit
parameters of the experimental MQC qubits \cite{Mooij2,Vion}. The
critical current in the QPC is close to the maximum
supercurrent of a single mode, $I_c\approx e\Delta/\hbar$ ($\sim$
400nA for Nb), and the Josephson energy, $E_J\approx\Delta$. Assuming
$L\sim $0.1nH, and $C\sim $ 0.1pF, we estimate
$\omega \sim 10^{11}$sec$^{-1}$,
and $E_L/\hbar\sim 10^{13}$sec$^{-1}\sim 10\Delta_{Nb}/\hbar$. For these
values, the $q_0$-factor is of order unity, and the reflectivity of the
contact must be small, $R \leq 0.01$, which is, in principle, accessible
in the experiments with atomic-size
QPC\cite{Goffman,Scheer}. However, even a slight increase of the
inductance will significantly decrease the $q_0$-factor, and the
constraint on the bare contact reflectivity will become less restrictive.
In summary, we estimate the upper bound for the qubit operation
frequency to be about $10^{10}$sec$^{-1}$.

The qubit operation frequency must significantly exceed the relaxation and dephasing
rates of the qubit. The relaxation and dephasing mechanisms (external 
flux fluctuations,
radiation, etc.), which have been extensively discussed for the MQC flux qubits
\cite{Makhlin2001}, are also relevant for the Andreev
level qubit. The interaction of the Andreev levels with microscopic degrees of 
freedom in the contact, primarily with the phonons, does not impose any further limitations.
Investigation of this problem has shown \cite{IvanovPhon} that the
relaxation rate is very sensitive to the Andreev level spacing: 
for small spacing, $\tau^{-1}\sim \tilde R^2\tau^{-1}_{ph}(\Delta)$ at small 
temperature, $T\ll \tilde E_a$. 
For $\tilde R < 0.01$ this relaxation rate
is smaller than the qubit operation frequency by at least a factor of
$10^{4}$.

We conclude with a discussion of the qubit-qubit interaction.
Let us consider an inductive coupling as the most relevant interaction for the flux qubits.
This coupling will introduce hybridization of the loop oscillators, which is
described by inserting the inductance matrix in the $H_{osc}$ term in 
Eq. (\ref{Hlin}). 
The averaging over flux fluctuations is conveniently performed using the oscillators normal modes.
The averaging procedure leads to the qubit Hamiltonians in Eq. (\ref{Hq}) 
with slightly different dressing factors, and to a Hamiltonian of direct qubit-qubit interaction.
For the two qubits  the effective interaction has the form,
$H_{int}=- (e/c\hbar)^2 ({\cal M} {\cal
I}_1 {\cal I}_2)\; \sigma_{z1}\sigma_{z2}$, where ${\cal M}$ is the mutual
inductance. The two-qubit configuration may consist, in particular, of a single QPC with two 
conducting modes; in this case ${\cal M}=L$.

In conclusion, we have developed a theory for the Andreev level qubit, 
a device consisting of a SQUID with a quantum point contact, 
combining the features of microscopic and macroscopic quantum systems. We 
derived the two-level Hamiltonian for the Andreev levels and showed that 
it is strongly dressed by the quantum fluctuations of the induced flux. 
We also derived the effective interaction Hamiltonian 
for inductively coupled qubits, and discussed the relevant circuit 
parameters for the qubit operation and read out. 

We acknowledge stimulating discussions with G. Johansson. 
This research was partially funded by project SQUBIT of the IST-FET
programme of the EC.

\end{document}